\documentclass[11pt]{article}

\usepackage{graphicx}
\usepackage{amssymb}
\usepackage{amsthm}
\usepackage{subfigure}
\usepackage{url}
\usepackage{amsmath}
 
\usepackage{anysize}
\marginsize{2.5cm}{2.5cm}{2.5cm}{2.5cm}

\begin{document}

\title{Towards Cytoskeleton Computers. A proposal\footnote{To be published as a chapter in the book Adamatzky A., Akl S., Sirakoulis G., Editors. From Parallel to Emergent Computing, CRC Press/Taylor \& Francis, 2019}.}

\author{
Andrew Adamatzky$^a$ \and
Jack Tuszynski$^b$ \and
J\"{o}rg Pieper$^c$ \and
Dan V. Nicolau$^d$ \and 
Rossalia Rinalndi$^e$ \and
Georgios Sirakoulis$^f$ \and
Victor Erokhin$^g$ \and
J\"{o}rg Schnau{\ss}$^{hi}$ \and
David M. Smith$^{hi}$\\
\cr
$^a$ University of the West of England, Bristol, UK\\
$^b$ Politecnico di Torino, Italy\\
$^c$ University of Tartu, Estonia\\
$^d$ McGill Univesity, Montr\'{e}al, Canada\\ 
$^e$ University of Salento, Italy\\
$^f$ Democritus University of Thrace, Xanthi, Greece\\
$^g$ CNR, University of Parma, Italy\\
$^h$ University of Leipzig, Germany\\
$^i$ Fraunhofer Institute for Cell Therapy and Immunology, Leipzig, Germany\\
}

\maketitle

\setcounter{tocdepth}{1}
%\tableofcontents

\section{Introduction}

Actin~\cite{holmes1990atomic} and tubulin~\cite{lowe2001refined} are key structural elements of Eukaryotes' cytoskeleton~\cite{cytoskeletonbook,Huber2013}. The networks of actin filaments (AF)~\cite{holmes1990atomic,lieleg2010structure,Golde2013} and tubulin microtubules (MT)~\cite{purich1984microtubule} are substrates for cells' motility and mechanics~\cite{fletcher2010cell,vicente2011cell,Golde2018}, intra-cellular transport~\cite{volkmann1999actin,schuh2011actin} and  cell-level learning ~\cite{hameroff1988coherence, rasmussen1990computational, ludin1993neuronal, conrad1996cross, tuszynski1998dielectric, priel2006dendritic, debanne2004information, priel2010neural, jaeken2007new,dayhoff1994cytoskeletal}. 
Ideas of information processing taking place on a cytoskeleton network, especially in neurons, have been proposed by Hameroff and Rasmussen in the late 1980s in their designs of tubulin microtubules automata~\cite{hameroff1989information} and a general framework of cytoskeleton automata as sub-cellular information processing networks~\cite{rasmussen1990computational,hameroff1990microtubule}.  Priel, Tuszynski and Cantiello discussed how information processing could be implemented in actin-tubulin networks of neuron dendrites~\cite{priel2006dendritic}. The hypothetical AF/MT information processing devices can transmit signals as travelling localised patterns of conformational changes~\cite{hameroff1982information,hameroff2002conduction}, orientational transitions of dipole moments~\cite{tuszynski1995ferroelectric,brown1997dipole,cifra2010electric}, and ionic waves~\cite{tuszynski2004ionic,priel2006ionic,sataric2009nonlinear}. While propagation of information along the cytoskeleton is well studied, in theoretical models, there are almost no experimental results on actual processing of information. Computational studies demonstrated that it is feasible to consider implementing Boolean gates on a single actin filament~\cite{siccardi2016boolean} and on an intersection of several actin filaments~\cite{siccardi2016logical} via collisions between solitons and using reservoir-computing-like approach to discover functions on a single actin unit~\cite{adamatzky2017logical} and filament~\cite{adamatzky2018discovering}.

We propose a road-map to experimental implementation of cytoskeleton-based computing devices. An overall concept is described in the following.
\begin{itemize}
\item Collision-based cytoskeleton computers  implement logical gates via interactions between travelling localisation (voltage solitons on AF/MT chains and AF/MT polymerisation wave fronts). 
\item Cytoskeleton networks are grown via programmable polymerisation. Data are fed into the AF/MT computing networks via electrical and optical means. \item Data signals are travelling localisations (solitons, conformational defects) at the network terminals. 
\item The computation is implemented via collisions between the localisations at structural gates (branching sites) of the AF/MT network. 
\item The results of the computation are recorded electrically and/or optically at the output terminals of the protein networks. 
\item As additional options, optical I/O elements are envisaged via direct excitation of the protein network and by coupling to fluorescent molecules. 
\end{itemize}

\section{Rationale behind our choice of cytoskeleton networks}

\subsection{Why AF and MT not DNA?} 

DNA is proven to act well as a nano-wire~\cite{warman1996dna,berlin2000dna,beratan1997dna}, however no transformations of signals have yet been observed. MT's show signal amplification~\cite{priel2006biopolymer}. AF/MT display very high-density charges (up to 105 e/micron) manifested by extensive changes in the electric dipole moment, the presence of an anomalous Donnan potential and non-linear electro-osmotic response to a weak osmotic stress~\cite{singh1984approach,lin1993novel, angelini2006counterions,raviv2007microtubule}. Charge density is approximately 40 times higher than that on DNA. As postulated by Lin and Cantiello~\cite{lin1993novel}, electrically forced ions are predicted to be entering one end of the AF/MT result in ions exiting the other (the ionic gradient develops along the filament). AF/MT are nonlinear inhomogeneous transmission lines supporting propagation of non-linear dispersing waves and localized waves in the form of solitons. AF/MT has a fundamental potential to reproduce itself via polymerisation. AF is a macro-molecular actuator~\cite{knoblauch2004biomimetic} and therefore actin-computing circuits could be embedded into molecular soft machinery~\cite{ariga2012evolution}. The polymerisation of AF/MT can be finely tuned and most desired architectures of computing circuits can be grown~\cite{schiff1979promotion,wear2000actin,hayes2006regulation}. Moreover, AF/MT supports propagating voltage solitons (Sect.~\ref{signalcarriers}). Also, both AFs and MTs form their own easily controllable networks, which is not the case with DNA. Furthermore, DNA is flexible mechanically and coiles up under various influences, while AFs and MTs are the two most mechanically rigid structures seen in cell biology~\cite{gittes1993flexural,sato1987dependence,mogilner1996cell,venier1994analysis,dima2008probing}. Rigid tube-like~\cite{schuldt2016tubes,glaser2016,Schnauss.2017Jove} or beam-like~\cite{sajfutdinow2018bricks} structures grown from small sets of DNA strands demonstrate single-filament and network properties similar to actin networks. %Rigid tube-like~\cite{schuldt2016tuning} or beam-like~\cite{sajfutdinow2018direct} structures grown from small sets of DNA strands demonstrate single-filament and network properties similar to actin networks

While the electrical transport along these types of structures have not yet been investigated, they would not be expected to show wholesale deviations from their underlying material.

\subsection{Why consider both AF and MT?}

While AF and MT are both protein filaments with quasi-linear geometries, their major difference is in both their biological functions and physical properties. Actin filaments are very thin (about 5~nm diameter) while MTs are very thick (25~nm diameter)~\cite{holmes1990atomic,lowe2001refined,Huber2013}.  MT are distinct in their dynamic instability, which includes both linear polymerisation and stochastically distributed catastrophes~\cite{bolterauer1999models,hinow2009continuous}. Also, at high concentrations they exhibit collective oscillations~\cite{houchmandzadeh1996collective}, which are unique in cell biology. MTs are in fact hollow cylinders with 1~nm pores on their surface allowing for ionic currents to flow in and out of the lumen. AF form branched networks in the presence of other proteins, while MTs form regular lattices with interconnections provided by MAPs. MTs have a much higher electrostatic charge per length than AFs and their C-termini carry 40\% of the charge, making them antenna-like objects.

\section{Carriers of information}
\label{signalcarriers}

The cytoskeleton protein networks propagate signals in the form of ionic solitons~\cite{tuszynski2004ionic, sataric2010solitonic,poznanski2017induced} and travelling conformation transformations~\cite{pokorny1997vibrations,pokorny2004excitation,friesen2015biological,mavromatos2017non,kavitha2014propagation} and breathers generated through electrical and mechanical vibrations~\cite{kavitha2017localized}.  Experiments with polarised bundles of AF/MT demonstrated that micro-structures when polarised can sustain solitary waves that propagate at a constant velocity without attenuation or distortion in the absence of synaptic transmission~\cite{poznanski2017induced}. We argue that the travelling localisations (solitons, defects, kink waves) transmit information along the cytoskeleton networks and that this information is processed/modified when the localisations interact/collide with each other. 

With regards to ionic waves we expect them to interact similarly to excitation waves in other spatially-extended non-linear media.  A thin layer Belousov-Zhabotinsky (BZ) medium is an ideal example.   A number of theoretical and experimental laboratory prototypes of BZ computing devices 
have been produced. They are image-processing and memory devices~\cite{kuhnert1986new, kuhnert1989image, kaminaga2006reaction},  
wave-based counters~\cite{gorecki2003chemical},
memory in BZ micro-emulsion \cite{kaminaga2006reaction},
neuromorphic architectures~\cite{ gorecki2006information, gorecki2009information, stovold2012simulating, gentili2012belousov, takigawa2011dendritic, stovold2012simulating,  gruenert2015understanding} and associative memory \cite{stovold2016reaction,stovold2017associative},
information coding with frequency of oscillations~\cite{gorecki2014information}, 
logical gates implemented in geometrically-constrained BZ medium~\cite{steinbock1996chemical, sielewiesiuk2001logical}, approximation of shortest path by excitation waves~\cite{steinbock1995navigating, rambidi2001chemical, adamatzky2002collision}, 
chemical diodes~\cite{DBLP:journals/ijuc/IgarashiG11},   and  other types of processors~\cite{DBLP:journals/ijuc/YoshikawaMIYIGG09, escuela2014symbol, gruenert2014understanding, gorecki2015chemical}. A range of prototypes of arithmetical circuits based on interaction of excitation wave-fronts has been implemented within the BZ methodology. These include Boolean gates~\cite{steinbock1996chemical, sielewiesiuk2001logical, adamatzky2004collision, adamatzky2007binary, toth2010simple, adamatzky2011towards}, including evolving gates~\cite{toth2009experimental} and   clocks~\cite{de2009implementation}.   A one-bit half-adder, based on a ballistic interaction of growing patterns~\cite{adamatzky2010slime}, was implemented in a geometrically-constrained light-sensitive BZ medium~\cite{costello2011towards}.  Models of multi-bit binary adder, decoder and comparator in BZ are proposed  in~\cite{sun2013multi, zhang2012towards, suncrossover, digitalcomparator}.

The cytoskeleton computer executes logical functions and arithmetical circuits via interaction of travelling localisations, i.e. via collision-based computing.

\section{Collision-based computing}

A collision-based, or dynamical, computation employs mobile compact finite patterns, mobile self-localised excitations or simply localisations, in an active non-linear medium. These localisations travel in space and perform computation when they collide with each other. Essentials of collision-based computing are described in the following~\cite{adamatzkyCBC}. 
Information values (e.g. truth values of logical variables) are given by either absence or presence of the localisations or other parameters of the localisations. 
The localisations travel in space and perfor computation when they collide with each other. 

 Almost any part of the medium space can be used as a wire, although, if a travelling localisation occupies the whole width of the polymer network, then AF/MT can be seen as quasi-one-dimensional conductors. 
Localisations can collide anywhere within a space sample; there are no fixed positions at which specific operations occur, nor location-specified gates with fixed operations. 
The localisations undergo transformations (e.g. change velocities), form bound states, annihilate or fuse when they interact with other mobile patterns. 
Information values of localisations are transformed as a  result of collision and thus a computation is implemented.

There are several sources of collision-based computing. Studies dealing with collisions of signals, travelling along discrete chains are only now beginning to be undertaken within the field of computer science. The ideas of colliding signals, which had been initiated in the nineteenth century physics and physiology, were then put in a context of finite state machines around 1965, when papers by Atrubin on multiplication in cellular automata~\cite{atrubin1965one}, Fisher  on generation of prime numbers in cellular automata~\cite{fischer1965generation}, and Waksman on the eight-state solution for a firing squad synchronisation problem were published~\cite{waksman1966optimum}. In 1982 Berlekamp, Conway and Gay~\cite{berlekamp1982winning} demonstrated that the Game of Life 2D cellular automaton (with just two cell states and eight cell neighbourhood) can imitate computing circuits. Gliders, small patterns of non-resting cell states, were selected as main carriers of information. Electric wires were mimicked by lines along which gliders travel, and logical gates were implemented via collisions of gliders (namely, Berlekamp, Conway and Gay employed annihilation of two colliding gliders to build a {\sc not} gate and combination of glider guns, generators of gliders, and eaters, structures destroying these mobile localisations, to implement {\sc and} and {\sc or} gates). Almost at the same time, Fredkin and Toffoli~\cite{fredkin1982conservative} have shown how to design a non-dissipative computer that conserves the physical quantities of logical signal encoding and information in a physical medium. They further developed these ideas in the conservative logic~\cite{fredkin1982conservative}, a new type of logic with reversible gates. The billiard-ball model (BBM) has been ingeniously implemented in 2D cellular automata by Margolus~\cite{margolus1984physics} with an 8-cell neighbourhood and binary cell states. This was later enriched with results of non-elastic collisions~\cite{margolus2002universal}, Turing universality of BBM~\cite{durand2002computing}, number conserving models and BBM on triangular latices~\cite{morita2002universal}. In 1986 Park, Steiglitz and Thurston~\cite{park1986soliton} designed a parity filter cellular automata (analogues of infinite response digital filters), which exhibit soliton-like dynamics of localisation. This led to the development of mathematical construction of a 1D particle machine, which performs computation by colliding particles in 1D cellular automata, and the concept of embedded computing in bulk media~\cite{squier1994programmable}. The constructions of particle machines and BBM are well complemented by recent advances in soliton dragging logic, acousto-optic devices, cascadable spatial soliton circuits and optical filters, see e.g.~\cite{blair2002gated}.

In computational experiments~\cite{siccardi2016logical} we demonstrated that it is possible to implement logical circuits by linking the protein chains. Boolean values are represented by localisations travelling along the filaments and computation is realised via collisions between localisations at the junctions between the chains. We have shown that {\sc and}, {\sc or} and {\sc not} gates can be implemented in such setups. These gates can be cascaded into hierarchical circuits, as we have shown on an example of {\sc nor}. The approach adopted has many limitations, which should be dealt with in further studies. The collision-based computing techniques could be free from their dependence on timing by adopting stochastic computing~\cite{gaines1969stochastic,alaghi2013survey} by converting the numbers to be processed into long streams of voltage solitons, or travelling defects, which represent random binary digits, where the probability of finding a `1' in any given position equals the encoded value. 

The ultimate goal here is to make general-purpose arithmetical chips and logical inference processors with AF/MT networks. We can achieve this as follows:
\begin{itemize}
\item  Single unit (globular actin) devices realise Boolean gates, pattern recognition primitives, memory encoded into limit cycles and attractors of global state transition graphs of the molecule. 
\item Collision-based logical gates are implemented in actin networks, cascades of gates are realised. 
\item The collision-based logical gates are assembled into arithmetic and logical units, 8-bit operations are implemented via collision of voltage solitons on actin networks with tailored architecture. 
\item Reversible gates (Fredkin and Toffoli gates) are realised via interactions of ionic waves. 
\item Logical inference machines and fuzzy controllers are produced from hybrid actin electronic components, electrical analog computing primitives realised in bundles of actin fibres, and memristor-based arithmetical units are prototyped.
\end{itemize}

\section{Memory}

\subsection{Meso-scale memory via re-orientation of filaments bundles}

When an AC electric field is applied across a small gap between two metal electrodes elevated
above a surface, rhodamine-phalloidin-labelled actin filaments are attracted to the gap and
became suspended between the two electrodes~\cite{arsenault2007confinement}. The filaments can be positioned at predetermined
locations with the aid of electric fields. The intensity of an electric field can be encoded
into amplitudes of the filaments' lateral 
fluctuations. Nicolau and colleagues demonstrated
the organisation of actin filamentous structures in electric fields, both parallel and perpendicular
to the field direction~\cite{hanson2005electrophoretic,ramsey2013electric,ramsey2014control}. This will act as memory write operation. To erase the info we
use a DC field, to align actin filaments transversely to the electric field. In the exploratory
part of this task, the memory device will be also studied in a context of low-power storage with
information processing capabilities
 
\subsection{Nano-scale memory via phosphorilation of polymer surface}
 
In a parallel set of computational experiments,  Craddock,  Tuszynski and Hameroff demonstrated that phosphorylation of the MT surface (Serine residues on tubulin's C-termini) mechanistically explains the function of calcium calmodulin kinase II (CaMKII) in neurons. This has been proposed to represent a memory code~\cite{hameroff2010memory,craddock2012cytoskeletal}. We will explore the ramifications of this hypothesis on upstream effects such as motor protein (e.g. kinesin and/or dynein) processivity and reorganization of the architecture of the neuronal cytoskeleton. Implementation of this memory code can lead to the development of molecular {\sc xor} as well as {\sc and} logical gates for signal processing within neurons. We will investigate how this can lead to complex functionality of the MT cytoskeleton in neurons and how this can be extended to the artificially manufactured hybrid protein-based devices.

\section{Interface}

\subsection{Optical I/O}

Optical I/O can be realised by direct excitation into the 350 nm absorption band of actin using a Nd:YAG laser~\cite{beavis1989matrix,malyarevich1998v,fan1987continuous}. Optical output requires the integration of hybrid systems of actin and fluorescent molecules or e.g. bacteriorhodopsin (BR)~\cite{henderson1990model}. BR will be used not only as an optical output system, but also as a light-induced proton pump~\cite{lozier1975bacteriorhodopsin}, allowing to vary a map of electrical potential distribution under illumination. 

To develop optical inputs one could excite polymer ensembles directly at 355~nm with a Nd:YAG laser using its third harmonics. An optical output could then be made by coupling actin strands with  BR and/or fluorescent markers~\cite{gite2000ultrasensitive}. Upon coupling with actin, the bacteriorhodopsin molecule will function as an emitter at approximately 700~nm~\cite{gudesen1999optical}. To increase intensity of the optical output one could use fluorescent markers. It is also worth exploring if BR could be used as a switchable input and output. Within the photo-cycle BR shifts from 568 to 410~nm in ms, but the time can be tuned by mutation. The broad 410~nm band is well overlapping the 350~nm actin band. This means that there will be uphill transfer at 20$^o$C, but also better overlap for downhill transfer depending on whether BR or actin is excited. One could also perform single molecule FRET experiments to monitor protein conformation changes~\cite{sherman2008using} and dynamics during signal propagation. It would be very useful to develop a system to measure and image voltage changes and propagation along individual polymer chains, similar to imaging voltage in neurons~\cite{peterka2011imaging}. By linking the voltage sensor to actin binding compounds or proteins one could then select an optimal sensor for imaging actin filaments. Also, it would be very important to link an actin voltage sensor to other proteins such as tropomyosin to cover the lattice of the polymers. As an alternative, we could use LifeAct~\cite{riedl2008lifeact}, which is a small peptide that interacts with actin filaments and may not inhibit the branching activity of the Arp2/3 complex since does not interfere with the interaction of many other actin binding proteins~\cite{riedl2008lifeact}. 

Single-molecule fluorescence methods such as F\"{o}rster Resonance Energy Transfer (FRET) are well suited for looking at molecular interactions and dynamics on the nano scale~\cite{crevenna2012effects}. The optimal approach to investigate the dynamics of the FRET signal depends on the time-scale of the dynamics. From the nano seconds to the micro seconds scale, one can use one of several approaches: either fluorescent correlation spectroscopy~\cite{thompson2002fluorescence} to visualise the anti-correlated signal of the donor and acceptor channels. 
As the anti-correlated signal is often masked by the positive correlation of other processes such as diffusion, it is necessary to be able to separate the FRET dynamics from other correlation signals. With the pulsed interleaved excitation technique~\cite{muller2005pulsed} we can perform a cross-correlation analysis in the presence and absence of FRET dynamics using the same data set. By then globally fitting the two cross-correlation curves, we can extract the specific FRET contribution. Another approach for investigating and quantifying the observed FRET dynamics is to use multi-parameter fluorescence detection (MFD)~\cite{weidtkamp2009multiparameter}. In MFD, the maximum amount of information is collected from each photon. This includes fluorescence wavelength, lifetime and anisotropy, allowing us to quantify FRET signals and resolve dynamics on the nanosecond time scale. Lastly, one can measure the nano-second correlation of a donor in the presence of an acceptor~\cite{nettels2007ultrafast}.

\subsection{Electrical I/O}

Electrical I/O could be implemented via multi-electrode array technology~\cite{spira2013multi}; this gives us time resolution on the order of milliseconds and offers the possibility to go towards device/chip configuration on top of which we can grow/deposit AT/MT bundles. These approaches might be augmented with nanofiber-light addressable potentiometric sensor~\cite{shaibani2016detection}] and stimulated-emission depletion microscopy (STED)~\cite{hell1994breaking,dyba2003immunofluorescence} combined with fluorescence correlation spectroscopy (FCS)~\cite{thompson2002fluorescence,ries2012fluorescence} with high spatial resolution (50-100~nm) and time resolution from tens of nanoseconds to milliseconds~\cite{lanzano2017measurement}. The external trigger signal is explored using a novel pump-probe approach, where protein dynamics initiated by any external signal are investigated in real time on microseconds to milliseconds timescales~\cite{pieper2008transient,pieper2010time}. This approach is capable of detecting the presence and timescale of soliton propagation, following electronic or optical excitation. Such localised conformation changes in proteins can be directly probed by neutron spectroscopy~\cite{pieper2009protein}. This conventional approach reveals a general mobility of the protein only and has already been successfully applied to globular and filamentous actin.  Electrical inputs as well as electromagnetic fields are known to affect cytoskeleton components in complex ways that include (de)polymerisation effects and ionic wave activation~\cite{fels2015fields}.

\subsection{Characterisation of travelling soliton-signals under dynamic conditions} 

\subsubsection{Millisecond time resolution}

Multi-electrode array (MEA) technologies~\cite{franke2012high} can be exploited to measure travelling ionic waves/voltage solitons (it must first be determined if the signal-to-noise ratio is sufficiently high to detect the information along the filaments). The spatial resolution depends on the density of microelectrodes in the arrays, and on the electronic components of the measuring circuit/amplifiers.

\subsubsection{Microseconds to tens of nanoseconds time resolution} 

The colour changes along the fibres can be monitored by means of fluorescence lifetime imaging (FLIM)~\cite{lakowicz1992fluorescence} coupled with super resolution optical microscopy based on stimulated emission depletion (STED)~\cite{hell1994breaking}, providing also spatial resolution on the order of 100~nm. F\"{o}rster resonant energy transfer (FRET)~\cite{van1994resonance, deniz1999single} can be also coupled to the FLIM-STED methods.
The FLIM-STED-FRET could be used with the aid of fluorescence voltage indicators~\cite{gong2015high,st2014high} or ratiometric optical sensors~\cite{park2003ratiometric}. Electrochromic dyes are ideally suited to monitor ``fast'' voltage changes, which is induced by the molecular Stark effect~\cite{kuhn2004high,kulkarni2017voltage}. Fluorescent quantum dots can be used as a more powerful alternative in this case, providing also multiplexing capabilities~\cite{park2012single,marshall2013optical}. In the case of ratiometric $\mu$-optical sensors, the ionic concentration on the surface of the polymers can be used to bind ratiometric ion-sensitive colloids~\cite{del2015ratiometric}, which are able to tune their emission as a function of ionic concentration and/or voltage (voltage solitons).  

\section{Growing cytoskeleton circuits}

\subsection{MT assembly}

MT assembly is well understood and can be controlled by experimental conditions of temperature, ionic concentrations and pH~\cite{schiff1979promotion,drechsel1992modulation,shelanski1973microtubule}. MTs can be (de)stabilised by various families of pharmacological agents~\cite{Kubitschke2017}. For example, taxane compounds are known to stabilise MTs while colchicine and its analogues are known to prevent MT formation~\cite{bollag1995epothilones}. Vinca alkaloids cap MTs preventing their continued polymerisation~\cite{calligaris2010microtubule}. By skilfully timing and dosing the administration of these compounds to the solution of tubulin in an appropriate buffer, assemblies of MTs can be generated with desirable length distributions. Moreover, using high concentrations of zinc added to these solutions~\cite{gaskin1977zinc}, various interesting geometrical structures can be generated in a controllable manner such as 2D sheets of tubulin with anti-parallel proto-filament orientations and macro-tubes which are cylinders made up of tubulin, whose diameters are approximately 10 times greater than those of MTs~\cite{downing1998tubulin}. Such structures can be useful in characterising capacitive, conductive and inductive properties of tubulin-based ionic conduction systems. Finally, interconnections between individual MTs can be easily created by mimicking natural solutions found in neurons, namely by adding microtubule-associated proteins (MAPs)~\cite{olmsted1986microtubule} to the MT-containing dishes. MAPs added there will form networks whose architecture can be determined by confocal microscopy and transmission electron microscopy (TEM) imaging experiments. In essence, there is an almost inexhaustible range of possible architectures that can be built on the basis of MT and MAP assemblies. Their conductive properties are at the moment a completely unexplored area of research, which, based on what we know about MT conductive properties, is potentially a treasure trove of ionic conduction circuitry. Combining these circuits with actin-based circuits leads to a combinatorial explosion of possibilities that can only be described as a revolutionary transformation in the field of bioelectronics.

\subsection{AF assembly}

The polymerisation and assembly of AF is well-understood and characterised and can be controlled by experimental conditions of temperature, ionic concentrations, pH and a variety of accessory proteins ~\cite{Huber2013}. To control the geometry of actin assembly we can adopt a micro pattern method~\cite{galland2013fabrication} where an actin nucleation promoting factor (NPF)~\cite{goley2004critical} is grafted to a surface in a well-defined geometry~\cite{letort2015geometrical}. 
Surface density and geometrical arrangement of NPFs on the surface can be closely controlled down to fewer than 10~nm using recent methods such as single molecule contact printing~\cite{sajfutdinow2017nanoscale}.

%Surface density and geometrical arrangement of NPFs on the surface could be closely controlled down to fewer than 10 nanometres using recent methods such as single molecule contact printing~\cite{sajfutdinow2017printing}. 
In the presence of a suitable mixture of proteins including the Arp2/3 complex ~\cite{mullins1998interaction}, this geometry of NPF will drive actin assembly and at the same time will impose specific boundary condition to allow self-organisation~\cite{galland2013fabrication}.  We propose to turn the permanent micro-patterns into dynamic and 3D micropatterns. To that end we will use the laser micro-patterning process based on protein coating with pulses of light. This method allows us to (1) perform contact-less micro-patterning, (2) control grafted protein density, (3) control micro-pattern geometry with a sub-micro-metric resolution, (4) design micro-patterns in 3D, (5) micro-pattern multiple proteins successively and (6) perform on-the-fly patterning and therefore on the fly actin assembly. We have already demonstrated how the polymerisation and/or the organisation of actin-based self-assembled `carpets' on surfaces can be controlled by the properties of the native buffer, and post-self-assembly/deposition~\cite{nicolau1999actin,mahanivong2002manipulation,alexeeva2005controlled}. It is possible to fabricate ordered patterns formed by AF, through the tuned interplay between F-actin self-assembly forces, and forces applied by the atomic force microscope (AFM) tip in a contact mode. More specifically, by increasing the force applied by the AFM tip we could observe the shift from the visualisation of individual actin filaments to parallel actin filaments ‘rafts’. Thus, we could produce ordered hybrid nano-structured surfaces through a mix-and-match nano-fabrication technology~\cite{naldi2010afm,naldi2009self}. It is also possible to induce bundled architectures of AF, either in linear geometries or with regularly spaced branched nodes by implementing counterion condensation~\cite{huber2012counterion}, depletion forces~\cite{schnauss2016a,schnauss2016b,glaser2016,strehle2017,huber2015depletion}, or by using natural protein-based~\cite{lieleg2010structure,Strehle2011} or synthetic DNA-peptide~\cite{lorenz2018crosslink} actin crosslinking molecules. Using these approaches, or experimentally favourable combinations, we are already able to self-assemble actin-based structures without any additional fabrication methods. Inducing a combination of actin-associated peptides and DNA-based template (as shown previously~\cite{lorenz2018synthetic}), self-assembling structures can be precisely biased to gain control over the systems architecture.

\section{Cytoskeleton electronics}

\subsection{Cytoskeleton containing electronic components}

Within this activity several actin-containing electronic components and their successive utilisation for the circuits of the computational systems could be realised. In particular, we focus on the following elements. A variable resistor will be realised as a pure actin layer or an actin layer alternated with conducting polymer  polyaniline in a Langmuir-Blodgett layer structure between two metal electrodes~\cite{erokhina2015polyaniline,erokhin2015method}. Variation in the conductivity can be determined by the state of actin and/or by the organisation of entire supra-molecular structure. In the case of photo-diodes several approaches can be attempted, such as realisation of structures where proteins can be interfaced with photo-isomerisable and/or photosensitive molecules. In the first case the photo-induced variation is likely due to the changes of the layer structure, while in the second case it can result from the variation of the carrier density. In the case of transistors, the starting structures can be based on PEDOT:PSS electro-chemical FET~\cite{berzina2009electrochemical,rivnay2018organic}. Actin in this case can be used whether as an additional material of the active channel, or as an inter-layer between the channel and electrolyte. In the first case, the conductivity variation can be due to the morphological conformation changes in the channel, while in the second case they can result from the variation of ionic permeability of the inter-layer. Schottky barrier element architecture is similar to the resistor configuration; however, the system is asymmetric --- different metals with significant difference in their work function can be used for contacting; the implementation can be tested with artificial conductive polymers~\cite{cifarelli2014non}. In the case of the capacitor, one can try planar and sandwich configurations: in the first case the capacitance variation can be due to the changing of dielectric properties of actin insulator in different states, while in the second case several effects could be responsible for it, such as thickness variation, redistribution of charges, resulting in the different conditions of the electric double layer formation. For all realised elements, variations of the properties with temperature can be studied making, thus, a basis of the thermistor realisation. The AF/MT containing electronic components can be used in experimental prototyping of resistor network for voltage summation, RC integrating network, RC differentiating network, summing amplifier. Feasibility of the cytoskeleton electronics can be evaluated in designs of variable function generators.

\subsection{Computing circuits with actin memristors}

Memristor (memory resistor) is a device whose resistance changes depending on the polarity and magnitude of a voltage applied to the device's terminals and the duration of this voltages application~\cite{chua1971memristor,strukov2008missing}. The memristor is a non-volatile memory because the specific resistance is retained until the application of another voltage. 
Organic memristive device~\cite{erokhin2005hybrid,erokhin2011thin} was developed for mimicking specific properties of biological synapses in electronic circuits~\cite{smerieri2008functional,smerieri2008polymeric}. It is adequate for the integration into systems with biological molecules due to its flexibility~\cite{erokhin2010bio} and biocompatibility~\cite{dimonte2015conductivity,tarabella2015hybrid}. The control of the conductivity state in this case can be done also by optical methods~\cite{pincella2011electrical,dimonte2015spectral,battistoni2016spectrophotometric}. A memristor implements a material implication of Boolean logic and thus any logical circuit can be constructed from memristors [6]. We can fabricate in laboratory experiments adaptive, self-organised networks of memristors based on coating actin networks with conducting polymers. Actin-based memristive circuits will be used to implement one-bit full adder~\cite{erokhin2012organic}, single-~\cite{demin2015hardware} and double-~\cite{emelyanov2016first} layer perceptrons and conditional learning circuits~\cite{erokhin2011material}.

\subsection{Logical inference machine}

The actin implication gates can be cascaded into a logical inference machine as follows. A Kirchhoff-Lukasiewicz (KLM) machine~\cite{mills2005extended,mills2006empty, mills2008nature} combines the power and intuitive appeal of analog computers with conventional digital circuits. The machine is made as actin sheet with array of probes interfaced with hardware implementation of Lukasiewcz logic arrays. The L-arrays are regular lattices of continuous state machines connected locally to each other. Each machine implements implication and negated implication. Arithmetic/logical functions are defined using implication and its negation. Array inputs are differences between two electrical currents. By discriminating values of input current differences, we represent continuous-value real analog, discrete, multiple-valued, and binary digital. Algebraic expressions are converted to L-implications by tree-pattern matching/minimisation.

\bibliographystyle{plain}
\bibliography{bibliography}

\end{document}